\documentclass[runningheads]{llncs}
\usepackage{graphicx}
\usepackage[colorinlistoftodos]{todonotes}

\begin{document}
\title{The L2L System for Second Language Learning Using Visualised Zoom Calls Among Students\thanks{The research was partly supported by Science Foundation Ireland under Grant Number SFI/12/RC/2289\_P2, co-funded by the European Regional Development Fund.}}
\titlerunning{The L2L System for Second Language Learning}

 \author{Aparajita Dey-Plissonneau$^1$\orcidID{0000-0003-1429-6861} \and \\Hyowon Lee$^2$\orcidID{0000-0003-4395-7702} \and \\Vincent Pradier$^3$\orcidID{0000-0002-7050-6408} 
 \and \\Michael Scriney$^2$\orcidID{0000-0001-6813-2630} \and \\Alan F. Smeaton$^2$\orcidID{0000-0003-1028-8389}}

\authorrunning{A. Dey-Plissonneau et al.}

%Paper:   68

\institute{$^1$School of Applied Languages and Intercultural Studies, \\
$^2$Insight Centre for Data Analytics, 
\\Dublin City University, Glasnevin, Dublin 9, Ireland\\
$^3$PSL University (Paris Sciences et Lettres)
60, rue Mazarine, 75006 Paris, France
\email{Alan.Smeaton@DCU.ie}}

\maketitle              % typeset the header of the contribution
\begin{abstract}
An important part of second language learning is conversation which is best practised with speakers whose native language is the language being learned. We facilitate this by pairing students from different countries learning each others' native language. Mixed groups of students have Zoom calls, half in one language and half in the other, in order to practice and improve their conversation skills.  We use Zoom  video recordings with audio transcripts  enabled which generates  recognised speech from which we extract  timestamped utterances and calculate and visualise conversation metrics on a dashboard.  A timeline highlights each utterance, colour coded per student, with links to the video in a playback window.  L2L was deployed for a semester and recorded almost 250 hours of zoom meetings. The conversation metrics visualised on the dashboard are a beneficial asset for both students and lecturers.

\end{abstract}

\section{Pedagogical / Technological background}

Virtual exchange programmes use technology to connect people from around the world for educational exchange. The emphasis is on international partnerships not only to build on content-knowledge but also to develop 21st century skills, global citizenship, intercultural understanding, empathy, and collaboration. This mission is  fulfilled by a number of  organisations \cite{apa_3}.

In foreign language learning and teaching, such synchronous and asynchronous virtual exchanges are commonly referred to as telecollaboration \cite{apa_2}. Telecollaboration facilitates the development of linguistic and interactional competencies in the foreign language, intercultural understanding, and the capacity to negotiate and collaborate in multicultural work environments. It has been integrated in some form  in foreign language pedagogy  since the arrival of the internet. More recently, high quality videoconferencing as a telecollaboration tool has afforded language learners a fast-paced synchronous interaction with native speakers of the target language. In response to the current Covid-19 crisis, telecollaboration efforts have been catapulted with more training, mentoring, webinars, and pre-mobility partnerships.

Facilitated by technological progress, innovative learning activities, such as students reviewing  recordings of their online interactions for learning and reflection, are gaining importance in language pedagogy in tertiary education \cite{apa_1replace}. While the potential of such telecollaboration-integrated curricula is huge, the possibilities of enhancing the telecollaboration experience by extending the affordances of videoconferencing is now  needed. Challenges such as lack of confidence in speaking in a second language, sustaining student engagement, creating student-led experiences, and a sense of community, all need to be addressed. This paper introduces the L2L system  developed to facilitate language students’ engagement through concrete visualisation of their synchronous participation in Zoom calls with native speakers. This  provides both students and lecturers with insightful feedback on the fast-paced synchronous interactions that take place while facilitating self- and peer-review in the post-session phase, thus making such fast-paced synchronous interactions pedagogically more meaningful. 

\section{Description of the prototype}

\begin{figure}[ht]
    \centering
    \includegraphics[width=0.75\textwidth]{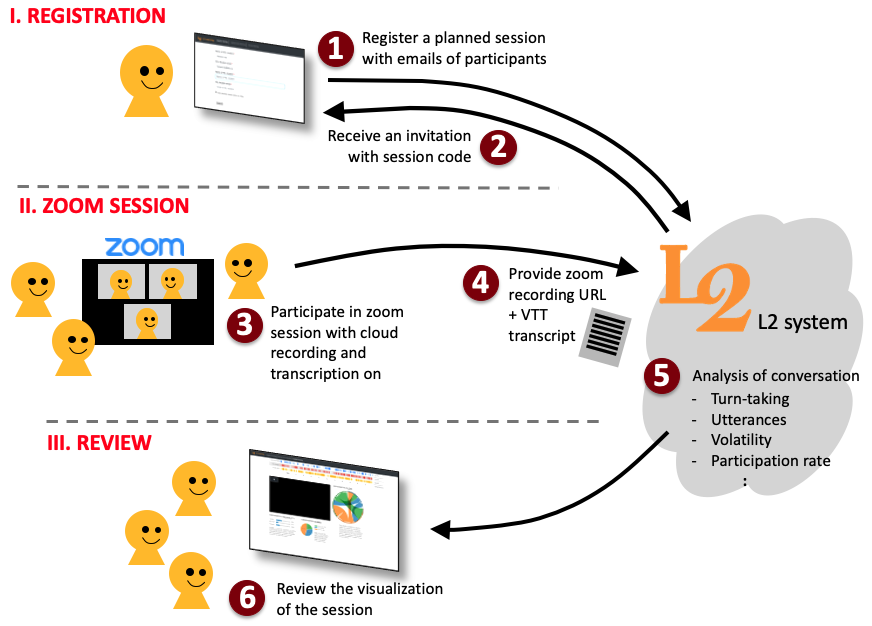}
    \caption{Using the L2L system - overall flow}
    \label{fig:flow-diagram}
\vspace{-8pt}
\end{figure}

The L2L system is based on the infrastructure of Help-Me-Watch, a system to provide personalised summaries of live video lectures \cite{lee2021attention}.
It makes use of Zoom's automatic audio transcription feature to analyse  conversations and to prepare the review page for each Zoom meeting. Figure~\ref{fig:flow-diagram} shows how the system works from the participating students' point of view. The reader should refer to the numbering in the Figure.

Before a Zoom conversation session starts, one of the participating students needs to register the session by visiting the L2L web interface and filling in the email addresses of all students who will join the session (1), which generates an immediate confirmation and invitation email with a unique Zoom session code (2) needed later to review the session.  Students then perform a Zoom-based conversation session with the Zoom cloud recording of the video and the audio transcript options enabled (3). Some time after the session ends, Zoom sends an automated email informing students/users of the availability of the cloud recording and other meta data including the audio transcript in VTT  format. One of the students then copies/pastes the URLs for the cloud recording of the video and the audio transcript  to the L2L system (4). Having received these, the L2L system analyses the session conversation by parsing the audio transcript text, calculating the characteristics of the session conversation in terms of turn-taking, utterances, volatility of dialogue, participation times of each student, etc. (5). The result of this analysis is then aligned with the cloud video recording of the session, and presented back to the students (6) as a web page. This setup is not straightforward but necessary in this version in order to anonymise students and secure the necessary data protection office and GDPR approval. Automation of this process and management by the system in forming and managing groups and meetings, is planned.

\begin{figure}[ht]
    \centering
    \setlength{\fboxsep}{0pt}
    \framebox{\includegraphics[width=0.95\textwidth]{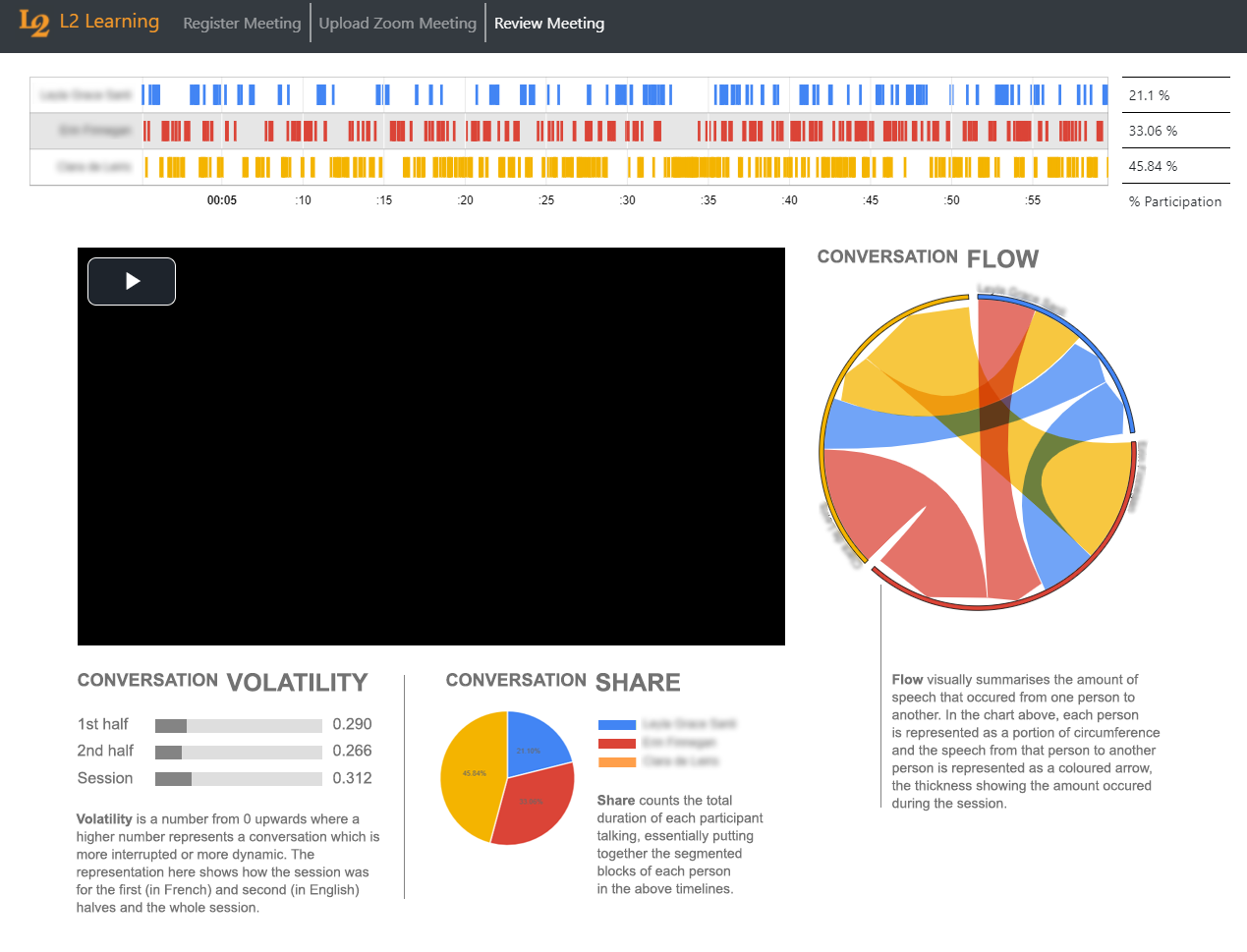}}
    \caption{Screenshot of review screen (student names are blurred)}
    \label{fig:L2-screen}
    \vspace{-9pt}
\end{figure}

The review screen shown in Figure~\ref{fig:L2-screen} summarises the analysis of s Zoom meeting.
On the top of the screen, a timeline  visualises the start and end of every utterance made by each student  during the session. Each  participating student shows his/her name with a  colour used throughout the graphics on the screen. The teacher or a student can click on any part of this timeline to immediately start playing the video in the window below  from the clicked point onwards. 
Conversation flow is a chord chart which visually summarises the amount of dialogue from one person to one another, thus indicating who was talking to whom. A pie chart shows the total duration of each participant’s talking. We also calculate conversational volatility, a measure usually applied to quantify the changing and dynamic nature of share prices on the stock market but when applied to conversational turn-taking it measures the dynamic nature of the dialogue.

\section{Use case to demonstrate  system relevance}

During  March-April 2021,  L2L  was used in  a telecollaborative project between  Dublin City University, an English-speaking University with 63 students involved and Paris Sciences et Lettres, a French-speaking University with 45 students involved.    The imbalance between the headcounts led us to organise groups into 45 Fr-Eng teams, 16  composed of three students (2 Eng and 1 Fr) and 29  composed of two students (1 Eng and 1 Fr). Some of the two-student teams were later turned into quadruplets, in order to train students in language interactions in the context of larger groups.  

The 45 teams collaborated for seven weeks on tasks focused on developing their interaction skills  in their target language, thanks to access to resources and  guidance provided by questions framing their discussions. As per the definition of the project, the discussions were managed half in French and half in English, so as to benefit from the mastery of the language by the native speaker, as a help to improve the learner’s skills. The topics of the  tasks changed every week and dealt with international news events and professional development content. For instance the students exchanged on international issues such as statelessness or professional questions such as job search and workplace inequalities between men and women. 

These frequent direct exchanges with native speakers of the target languages were opportunities for  students to increase the quantity of input in a wider variety of accents, but also the amount of moments of interaction practice in their target language. It provided  students with a unique environment in which they could face the difficulties of exchanging in a foreign language, in the reassuring context of a team whose members could help them develop their skills.

\section{Results and outcomes achieved}
The L2L system has been used for the analysis of almost 250 student Zoom meetings for second language  learning.  Student feedback on how the visualisations helped their reflection on meetings by allowing them to see overviews and replay specific parts, is described elsewhere \cite{eurocall} but we can summarise this by highlighting that 90\% said L2L helped their conversational language learning and 77\% said it helped their reflection.

At this point the system is operational and quite easy to use in terms of Zoom recordings and sharing Zoom cloud files but there were difficulties for some students in initially setting it up. These were mostly around configuring the correct Zoom settings and enabling the correct type of video recording and audio transcription on Zoom.  Our instructions to students on the steps to achieve this were by necessity  long in order to preserve their anonymity and for the  future, subject to DPO approval,  we intend to de-anonymise student identities and extend the visualisation dashboard to allow week-on-week progress for each student to be seen.

\section{Future agenda and next steps}

L2L has been used for hundreds of hours of student meetings  in an English-speaking country learning French, collaborating with students in a French-speaking country learning English. Following some initial teething problems with Zoom configurations, the system operates without problem.
When Zoom generates an audio transcript of the spoken dialogue, it is configured to recognise English, so when the speakers speak French the actual transcribed text is unrecognisable. 
Because our system only uses the timing information (who speaks when, and for how long) this is not an issue for our analysis. This means that  L2L  can be used for any language pair and broadening the languages is one of  our next steps.

%In case your paper is accepted, you would have to prepare a one-page document to describe technological aspects of your tool, along with its relationship with the conference topic. This document will not be published in the proceedings, but it will be needed for organizing and delivering the demonstration sessions accordingly.

%For citations of references, we prefer the use of square brackets
%and consecutive numbers. 

%Citations using labels or the author/year
%convention are also acceptable. The following bibliography provides
%a sample reference list with entries for journal
%articles~\cite{ref_article1}, an LNCS chapter~\cite{ref_lncs1}, a
%book~\cite{ref_book1}, proceedings without editors~\cite{ref_proc1},
%and a homepage~\cite{ref_url1}. Multiple citations are grouped
%\cite{ref_article1,ref_lncs1,ref_book1},
%\cite{ref_article1,ref_book1,ref_proc1,ref_url1}.

\bibliographystyle{splncs04}
\bibliography{L2L-ectel}

%Author, F.: Article title. Journal \textbf{2}(5), 99--110 (2016)
%Author, F., Author, S.: Title of a proceedings paper. In: Editor,
%F., Editor, S. (eds.) CONFERENCE 2016, LNCS, vol. 9999, pp. 1--13.
%Springer, Heidelberg (2016). \doi{10.10007/1234567890}
%Author, F., Author, S., Author, T.: Book title. 2nd edn. Publisher,
%Author, A.-B.: Contribution title. In: 9th International Proceedings
%on Proceedings, pp. 1--2. Publisher, Location (2010)
%LNCS Homepage, \url{http://www.springer.com/lncs}. Last accessed 4
%\end{thebibliography}
\end{document}